\documentclass[a4paper,12pt]{article}
\usepackage{amsmath}
\usepackage{graphicx}

\begin{document}

\begin{center}

{\Large \bf Constraints on composite Dirac neutrinos from
observations of galaxy clusters}\\[20mm]

R. S. Hundi\footnote{E-mail address: tprsh@iacs.res.in} and
Sourov Roy\footnote{E-mail address: tpsr@iacs.res.in}\\
Department of Theoretical Physics,\\
Indian Association for the Cultivation of Science,\\
2A $\&$ 2B Raja S.C. Mullick Road,\\
Kolkata - 700 032, India.\\[20mm]

\end{center}

\begin{abstract}
Recently, to explain the origin of neutrino masses a model
based on confining some hidden fermionic bound states into right-handed
chiral neutrinos has been proposed. One of the consequences
of condensing the hidden sector fields in this model is
the presence of sterile composite Dirac neutrinos of keV mass, which can form
viable warm dark matter particles. We have analyzed
constraints on this model from the observations of satellite based
telescopes to detect the sterile neutrinos in clusters of
galaxies.

\end{abstract}


\newpage

\section{Introduction}
\label{S:intro}

Since the discovery of neutrino oscillations
in the solar \cite{solar} and atmospheric \cite{atmos}
neutrino experiments, neutrinos
have played a vital role in the extension
of standard model. Although gauge hierarchy problem is the major
motivation for physics beyond the standard model, the
tiny masses of neutrinos can guide the model building part
of new physics in the neutrino sector. Theoretically,
the neutrinos can be either Dirac or Majorana fields,
depending on if lepton number is conserved or not in nature.
In the literature, models which explain the
tiny neutrino mass scale of 0.1 eV are mostly based on the
seesaw mechanism \cite{seesaw} which requires lepton number violating
Majorana neutrinos. A test for the Majorana neutrinos
is the existence of neutrinoless double beta decay process,
which has not been found in the experiments conducted so far.
In the light of this, models based on Dirac neutrinos have
also been proposed \cite{DirNeu}. Recently, in a model known as composite
Dirac neutrinos \cite{GR}, Dirac neutrino
masses have been proposed by conceiving right-handed neutrinos
as composed objects of hidden fermionic chiral bound states at a high scale $\Lambda$
\cite{AG,dCDN}. The hidden chiral bound states and standard model fields arise
due to confinement of an ultraviolet preonic theory at
another higher scale $M\gg\Lambda$. The gauge
symmetries of the hidden sector and the standard model can
be broken in such a way that in the low energy regime a gauge
symmetry is left unbroken which could be equivalent to the
gauged U(1)$_{\rm B-L}$, and hence in this model only Dirac
neutrino masses arise. The neutrino masses in this model come out
to be tiny due to the suppression factor of $\Lambda/M$.
One of the phenomenological consequences of this model is that
at the scale $\Lambda$, apart from chiral right-handed neutrinos
non-chiral sterile neutrinos can be produced due to confinement.
The mass scale $m_s$ of these sterile neutrinos depend on the nature
of confinement, and thus on both $M$ and $\Lambda$.
In a particular case and for suitable values of $M$ and $\Lambda$
its mass scale is $m_s\sim$ keV, which is the right amount to form
a warm dark matter particle \cite{wdm}.
These sterile states
have small mixing with active neutrinos, which is roughly
$\theta\approx m_\nu/m_s$, $m_\nu$ being the light active neutrino mass scale.

The existence of dark matter in the universe is well established
by galactic rotation curves, cosmic microwave background
radiation and gravitational lensing. The nature of dark matter
is unknown, yet about 20$\%$ of the universe is filled with it.
The issue of dark matter is another compelling reason for the
extension of standard model. Sterile neutrinos, which are
gauge singlets, have been thought to be good warm dark
matter particles \cite{wdm}. These
fields exist in many extensions of the standard model, one of
them is the composite Dirac neutrino model which is described
in the previous paragraph. Provided the mixing angle $\theta$
between sterile and active light neutrinos is sufficiently small,
the leading decay life time for keV mass sterile states into
three active neutrinos
can be larger than the age of the universe, thus making
them perfect dark matter candidates. One of the major channels
to detect a sterile neutrino is its one-loop decay into photon
and an active neutrino. For a keV mass sterile neutrino the emitted
photon energy would be in the X-ray region, and this decay feature
can be seen in the diffuse X-ray background of the universe or
in the X-ray flux from a cluster of galaxies. There would be
continuum X-ray background from a cluster of galaxies, nevertheless,
the signal from a photon due to sterile neutrino decay should be
a sharp peak on top of the X-ray background.

Satellite based experiments like {\it Chandra}, {\it XMM-Newton}, etc
have been operated to detect the decay line feature of photon due to
sterile neutrinos in the clusters of galaxies and also in the
diffuse X-ray background. The null results of this search put
exclusion area in the plane of sterile neutrino mass $m_s$
and the mixing angle $\theta$ \cite{stercon,AFT,XRB,BNRS,Lyman}.
As described above, in the
composite Dirac neutrino model the sterile neutrino mass
and the mixing angle are related to the confinement
scales $M$ and $\Lambda$ in such a way to provide a
viable warm dark matter particle and also a natural mechanism
for neutrino masses. The negative results of the
above mentioned experiments on the sterile neutrino decay
can put constraints on the composite scales $M$ and
$\Lambda$. In the original paper \cite{GR}, order of
magnitude estimations have been made on the model parameters
to explain the neutrino masses and dark matter.
In this work we study quantitative constraints
on the parameters $(\Lambda,M)$ of the composite Dirac
neutrino model due to these various X-ray based experiments.

The paper has been organized as follows. In the next section
we give a brief overview on the composite Dirac neutrino
model. In Sec. \ref{S:prob} we describe about some X-ray based experiments
and their findings. We apply these results in the specified composite
Dirac neutrino model and present the constraints on the
model parameters $(\Lambda,M)$. We conclude in
Sec. \ref{S:con}.

\section{Composite Dirac Neutrino model}
\label{S:mod}

This model assumes an ultraviolet (UV) preonic theory which confines
into standard model and hidden sector fields at a high scale $M$
\cite{GR,AG}. The symmetry of this theory spontaneously
breaks into $G_c\otimes G_F\otimes G_{\rm SM}$ at the scale $M$. Here
$G_c$, $G_F$ and $G_{\rm SM}$ are confinement, flavor and standard model
gauge symmetries, respectively. Below the scale $M$ the necessary fields
in our context are the following: (a) standard model lepton doublets
$L$ which are $G_c\otimes G_F$ singlets; (b) hidden fermionic chiral bound states $q$,
which are singlets under $G_{\rm SM}$ but charged under $G_c\otimes G_F$;
(c) scalar condensate $\phi$ which transforms as doublet under the electroweak
group of $G_{\rm SM}$, charged under $G_F$ and singlet under $G_c$.
The condensate $\phi$ can be interpreted as Higgs doublet of the standard
model. Now, consider a combination of $n$ hidden chiral bound states as $q^n$.
Imagine that the charges of $\phi$ and $q^n$ are arranged in such
a way that $\phi^* q^n$ is a singlet under $G_c\otimes G_F$. Then,
in the effective field theory below the scale $M$ there can be an
irrelevant operator in the Lagrangian of the form 
\begin{equation}
{\cal L} = \frac{\lambda}{M^{3(n-1)/2}}\bar{L}\tilde{\phi}q^n,
\label{E:beM}
\end{equation}
where $\tilde{\phi}=-i\sigma_2\phi^*$, $\sigma_i$ are the Pauli
matrices.
The hidden sector of this model can undergo one more confinement
at a scale $\Lambda$ much below the scale $M$, thereby condensing
the $q^n$ into a right-handed bound state $N_R$ as
\begin{equation}
q^n\to N_R\Lambda^{3(n-1)/2}.
\label{E:beL}
\end{equation}
Since the bound states are fermionic particles, $n$ should be odd and $n\geq 3$.
Plugging eq. (\ref{E:beL}) into eq. (\ref{E:beM}), in the low
energy regime one gets
\begin{equation}
{\cal L} = \lambda \left(\frac{\Lambda}{M}\right)^{3(n-1)/2}
\bar{L}\tilde{\phi}N_R,
\label{E:numt}
\end{equation}
The compound field $q^n$ is supposed to contain a right-handed
spin-1/2 Lorentz representation, and hence the field $N_R$, which
forms due to condensation below $\Lambda$, can be interpreted
as chiral right-handed neutrino. We get Dirac neutrino masses
for neutrinos from the above equation after electroweak symmetry
breaking. The mass scale
of these neutrinos can be of order 0.1 eV
by appropriately choosing the suppression factor $\epsilon =
\frac{\Lambda}{M}$ for $\lambda\sim {\cal O}(1)$. Let us
note in passing that composite Majorana neutrinos have also been
studied with a mini-seesaw mechanism for neutrino masses
\cite{McD}.

In the previous paragraph we have given the essential idea of
explaining the smallness of neutrino masses by conceiving
all the standard model fields in the low energy theory arising
as condensates of high energy UV preonic theory.
An important point to note here is that the masses of sterile neutrinos
depend on the type of interactions of
the fermionic bounds states, from which they are formed.
If these fermionic bound states are gauge singlets interacting with
scalar condensates via heavy gauge bosons
or massive scalars at the energy scale $M$, then the sterile neutrinos acquire
masses at loop level and their mass scale would be suppressed
to $m_s\sim\frac{\Lambda^3}{M^2}$. This procedure has been dubbed as secondary
mass generation mechanism \cite{GR,2mass}.

At the confinement scale $\Lambda$
the symmetry $G_c\otimes G_F$ breaks into $G^\prime_c\otimes G^\prime_F$.
All chiral bound states are $G^\prime_c$ singlets but transforms
under $G^\prime_F$. A simple choice for the $G^\prime_F$ is
$U(1)_F$. The charges of the fields $L$, $\phi$ and $N_R$ can
be chosen under $U(1)_F$ in such a way that after the
electroweak symmetry breaking a gauge symmetry $U(1)_a$,
which is an axial combination of the hypercharge group
$U(1)_Y$ and $U(1)_F$, is left unbroken. It can be shown
that the symmetry $U(1)_a$ is equivalent to lepton
number \cite{GR}. Hence, in this model lepton number is conserved
and only Dirac neutrino mass terms exist.

A detailed model by including mass terms for quarks and leptons
has been constructed in \cite{GR} with the choice $G^\prime_F$ =
$U(1)_F$. It has been shown that the axial combination of
$U(1)_Y$ and $U(1)_F$ would be isomorphic to $U(1)_{B-L}$, which
remains as an exact symmetry after electroweak symmetry breaking.
Moreover, the charges assigned under $U(1)_F$ satisfy requirements
for the cancellation of anomalies in this model. In general, at
the confinement scale $\Lambda$ some number of right-handed
bound states $N_R^I$, $I=1,\cdots,N$, can be produced. Out of which,
$I=i=1,2,3$ should
be chiral right-handed neutrinos in order to give masses to
three neutrinos through mass terms of the form of eq. (\ref{E:numt}).
The remaining $I=\alpha=4,\cdots,N$ can form
vector-like singlet neutrinos, for which the left-handed components
$N_L^\alpha$ exist at the confinement scale $\Lambda$. The
fields $N_L^\alpha,N_R^\alpha$ are by nature sterile and form
Dirac fermions. The mass scale of these sterile neutrinos could
be $\Lambda$. However, as
explained before, in the event of a secondary mass generation
\cite{GR,2mass} their masses would be suppressed by an
additional factor $\epsilon^2$. We consider the possibility of
secondary mass generation mechanism, since only in this case
we can conceive sterile neutrinos as keV warm dark matter particles
and also satisfy bounds due to big-bang nucleosynthesis \cite{dCDN}. 
The general mass terms in the neutrino sector in the low energy
regime can be written as follows.
\begin{equation}
{\cal L} = \lambda_{iI}\epsilon^{3(n_I-1)/2}
\bar{L}^i\tilde{\phi}N_R^I + \Lambda\epsilon^2\bar{N}_L^\alpha d_\alpha N_R^\alpha
+ {\rm h.c.},
\end{equation}
where the elements of $\lambda,d$ are ${\cal O}(1)$ constants,
and the repeated indices in $i,I,\alpha$ should be summed over.
The suppression in $\epsilon$ can be taken out as an overall constant
in the first term of the above equation by redefining
$\epsilon^{3(n_I-1)/2}
\lambda_{iI}\equiv \epsilon^{\tilde{n}}\tilde{\lambda}_{iI}$,
where $\tilde{n}$ = min$_I[3(n_I-1)/2]$. After doing this, the
mass terms in the basis of active and sterile neutrinos are
\begin{equation}
{\cal L}_m = \Lambda \left(\bar{N}^i_L~~\bar{N}^\alpha_L\right) A
\left(\begin{array}{c}
N^i_R \\
N^\alpha_R
\end{array}\right) + {\rm h.c.},
\end{equation}
where
\begin{equation}
A = \left(\begin{array}{cc}
\delta\tilde{\lambda}_{3\times 3} & \delta\tilde{\lambda}_{3\times k} \\
0 & \epsilon^2d_{k\times k}
\end{array}\right),\quad
\delta = \frac{v}{\Lambda}\epsilon^{\tilde{n}}.
\label{E:At}
\end{equation}
Here, $\langle \phi^0\rangle = v$ is the vacuum expectation value
of the Higgs field. In eq. (\ref{E:At}), the dimensions of various
matrices of $\tilde{\lambda}$ are indicated with $k=N-3$ and
$d_{k\times k}$ is a diagonal matrix of dimension $k$. The off-diagonal
elements in the matrix $A$ of the above equation give mixing
between the active and sterile neutrinos. In order to see the
mass eigenvalues of active and sterile neutrinos, consider the
simple case of $\tilde{n}$ = 3. Then, the mass eigenvalues and
the mixing angle $\theta$ between the active and sterile
neutrino states are
\begin{equation}
m_\nu\sim v\epsilon^3,\quad m_s\sim \Lambda\epsilon^2,\quad
\theta\sim \frac{v}{\Lambda}\epsilon.
\label{E:masnmix}
\end{equation}
Here, we have neglected the ${\cal O}(1)$ constants of $\tilde{\lambda}$
and $d$ elements. The mixing angle between the active and sterile neutrinos
has come out to be a ratio of their respective masses. To fit the neutrino
oscillation data in this model, we have to arrange the dimensionless
parameters $\tilde{\lambda}_{3\times 3}$ accordingly. Unlike the charged
lepton masses, there is a mild hierarchy in the neutrino mass eigenvalues.
Hence, the form for $m_\nu$ in the above equation gives a rough scale
for the neutrino masses. Now, consider
$\Lambda\sim$ TeV which satisfies the big-bang nucleosynthesis bounds
\cite{dCDN}. A neutrino mass scale of 0.1 eV can be fitted for
$M\sim 10^4$ TeV. For this set of $\Lambda$ and $M$ values the sterile
neutrino mass scale would be around keV and the mixing between the active
and sterile states is $\sim 10^{-5}$. The keV mass sterile neutrino
with a mixing of $\sim 10^{-5}$ with active neutrino can decay at
tree level through $Z$ boson into three active neutrinos. This decay
channel is the leading one and it gives a life time which is
considerably larger than the current age of the universe \cite{wdm}.
Thus, these keV sterile neutrinos are perfect warm dark matter particles.
Also, in this model, the mixing
with active neutrinos allows the sterile states to decay radiatively
into a neutrino and a photon. As explained in Sec. \ref{S:intro},
this decay channel has been probed in the X-ray based experiments
and constraints have been put in on the sterile neutrino parameters
\cite{stercon,AFT,XRB,BNRS,Lyman}.
Moreover, in this model, the mass scale of the sterile neutrinos
and their mixing angle with the
active neutrinos are in the right ball park region of the
analysis done in the X-ray based experiments.

In the next section we briefly describe about some of the techniques
in probing the decay of sterile neutrinos with X-ray telescopes
and their findings. These techniques can be employed in the composite
Dirac neutrinos and we can exclude some parametric space of
the model due to negative results in the experimental findings. Below
we describe the search strategies for the sterile neutrino flux
from the clusters of galaxies and analyze them in the composite
Dirac neutrino model. However, there are other ways to
probe sterile neutrinos in astrophysical experiments such as
diffuse X-ray background
spectrum analysis of the universe and the Lyman-$\alpha$ forest observations.
The studies on the diffuse X-ray spectrum of the universe give an
upper bound \cite{AFT,XRB}
and the observation from Lyman-$\alpha$ forest gives a lower bound
\cite{Lyman}
on the sterile neutrino mass. It is worth to consider these studies,
however, in this work we are not analyzing restrictions due to
these on the considered model.

\section{Probing sterile neutrinos in X-ray telescopes}
\label{S:prob}

It is believed that dark matter is clumped in clusters of
galaxies. If these clusters of galaxies contain keV sterile neutrinos
as warm dark matter candidate, we should detect photon flux in
the X-ray regime due to sterile neutrino decay, in the direction
of the cluster. However, the signal due to the sterile neutrino
decay has a strong background from the continuum emission of
X-rays due to the intra-cluster gas of the cluster of galaxies.
In the case that the signal is stronger than the background,
a sharp peak due to the decay of sterile neutrino should be
seen on top of the continuous X-ray background. The position of
the peak gives half the mass of the sterile neutrino. Experiments
such as {\it Chandra}, {\it XMM-Newton}, etc have been launched to detect
the X-ray spectrum from various clusters of galaxies. An analysis
done on the Willman 1 cluster has given an indication of the
existence of sterile neutrino of mass 5 keV \cite{W1}. However, this
result needs to be confirmed by others. Here, we describe the
analysis done on the Virgo and Coma clusters from the data collected
by {\it Chandra} and {\it XMM-Newton}, where a sharp
peak due to the sterile neutrino decay has not been found in the
X-ray spectrum. This negative search
gives some exclusion area in the sterile neutrino parameters.

In a cluster where dark matter decays into photons, the flux
from it as observed on the earth is
\begin{equation}
F=\frac{\cal L}{4\pi D_L^2},
\label{E:F}
\end{equation}
where $D_L$ is the luminosity distance of the cluster from
the earth and ${\cal L}$ is the luminosity of the source
which is given by
\begin{equation}
{\cal L} = \frac{E_\gamma}{m_s}M_{\rm DM}^{\rm fov}\Gamma_\gamma.
\label{E:lum}
\end{equation}
Here, $E_\gamma$, which in this case is $m_s/2$, is the energy of
the emitted photon and
$M_{\rm DM}^{\rm fov}$ is the mass of the dark matter in the cluster
as observed in the field of view of the telescope on the earth.
$\Gamma_\gamma$ is the decay width of the sterile neutrino into
a photon and an active neutrino \cite{PW}. The decay width
$\Gamma_\gamma$ depends on if the neutrinos are Dirac or Majorana.
However, in the case of Dirac neutrinos the value of $\Gamma_\gamma$
would be half the times that of the corresponding decay
width for the Majorana neutrinos \cite{PW}. As explained
previously, the decay feature of sterile states in the
X-ray spectrum is not observed by {\it Chandra} and {\it XMM-Newton}
telescopes. This put constraint on the flux due to sterile
neutrino decay, eqs. (\ref{E:F}) and (\ref{E:lum}),
that it should be less than the
observed flux. Based on this idea, we now describe the analysis
done by two different groups \cite{AFT,BNRS}, where constraints
on sterile neutrinos have been obtained. In both the analysis
that we describe below \cite{AFT,BNRS}, Majorana neutrinos have been assumed.
In order to translate these constraints into the case of
composite Dirac neutrino model, where only Dirac neutrinos
exist, we appropriately fix the factor 2 in the flux relation
and present our results.

In \cite{AFT}, the data from Virgo cluster by {\it Chandra} telescope
has been analyzed. It is claimed that {\it Chandra} telescope has
a background signal of 2$\times 10^{-2}$ counts s$^{-1}$ in
a 200 eV energy bin. It has been estimated that to overcome
this background at 4$\sigma$ level, a signal of flux $10^{-13}$
erg cm$^{-2}$ s$^{-1}$ should be detectable by the {\it Chandra}
telescope with an integration time of 36000 s.
After plugging the decay width $\Gamma_\gamma$ \cite{PW}
in eq. (\ref{E:lum}), the flux from a cluster can be written as
\begin{equation}
F\approx 5.1\times 10^{-18}~{\rm erg~ cm^{-2}~ s^{-1}}\left(
\frac{D_L}{\rm 1Mpc}\right)^{-2}\left(\frac{M^{\rm fov}_{\rm DM}}{10^{11}M_\odot}
\right)\left(\frac{\sin^22\theta}{10^{-10}}\right)
\left(\frac{m_s}{1{\rm keV}}\right)^5.
\label{E:aftF}
\end{equation}
The values for $D_L$ and $M^{\rm fov}_{\rm DM}$ for Virgo cluster
are given in \cite{AFT}. A stringent upper limit on $m_s$ can
be achieved if we assume that sterile neutrinos make up 100$\%$
of the dark matter of the universe. An approximate formula for
the relic density of sterile neutrinos is \cite{AFP}
\begin{equation}
\Omega_sh^2\approx 0.3\left(\frac{\sin^22\theta}{10^{-10}}\right)
\left(\frac{m_s}{100~{\rm keV}}\right)^2,
\label{E:rds}
\end{equation}
where $\Omega_s$ is the ratio of density of sterile neutrinos to
the total density of the universe and the current value
of $h$ is 0.72. Putting $\Omega_s$ = 0.3 in eq. (\ref{E:rds}),
which is roughly the current value for the relic abundance of the universe,
we can eliminate
the mixing angle in eq. (\ref{E:aftF}). Then, demanding that the
flux due to dark matter from the Virgo cluster to be less than
the minimal detectable flux
of $10^{-13}$ erg cm$^{-2}$ s$^{-1}$, an upper bound of 5 keV on the
sterile neutrino mass has been estimated \cite{AFT}.

The analysis done on the data collected by {\it Chandra} telescope
from Virgo cluster is more general \cite{AFT}. We apply this
analysis on the composite Dirac neutrino model. As explained
in Sec. \ref{S:mod}, in principle in this model there can be
arbitrary number of sterile neutrinos. However, without loss of generality,
we can take the mass $m_s$ of the lowest sterile neutrino and
its mixing angle $\theta$ with the active neutrinos to be
the same as in eq. (\ref{E:masnmix}).
The general analysis done on the data collected by the {\it Chandra}
telescope, if applied on this model, put constraints on the
model parameters $\Lambda$ and $M$, which are the confinement
scales at two different energies of the theory. However,
as explained before, we have to put a half-factor in the flux
relation of eq. (\ref{E:aftF}), since the neutrinos have
Dirac nature in this model. Putting
this half-factor and using eq. (\ref{E:masnmix}), the relations
for the X-ray flux and relic abundance of sterile neutrino
will take the following form, respectively.
\begin{equation}
F\approx 10.2\times 10^{-11}~{\rm erg~ cm^{-2}~ s^{-1}}\left(
\frac{D_L}{\rm 1Mpc}\right)^{-2}\left(\frac{M^{\rm fov}_{\rm DM}}{10^{11}M_\odot}
\right)\left(\frac{v}{\rm 1TeV}\right)^2 \left(\frac{\Lambda}{\rm 1TeV}\right)^{15}
\left(\frac{10^4~{\rm TeV}}{M}\right)^{12},
\label{E:cdmF}
\end{equation}
\begin{equation}
\frac{\Omega_sh^2}{0.3}\approx 4\times\left(\frac{v}{\rm 1TeV}\right)^2
\left(\frac{\Lambda}{\rm 1TeV}\right)^6
\left(\frac{10^4~{\rm TeV}}{M}\right)^6.
\label{E:cdmRD}
\end{equation}

\begin{figure}
\begin{center}
\includegraphics[height=2.5in,width=2.5in]{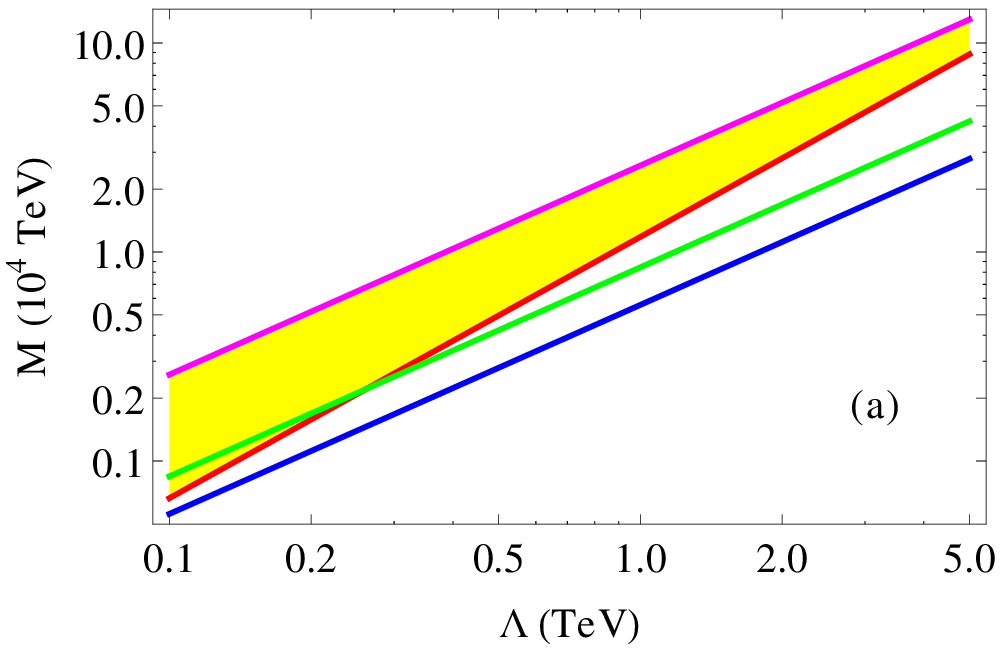}

\includegraphics[height=2.5in,width=2.5in]{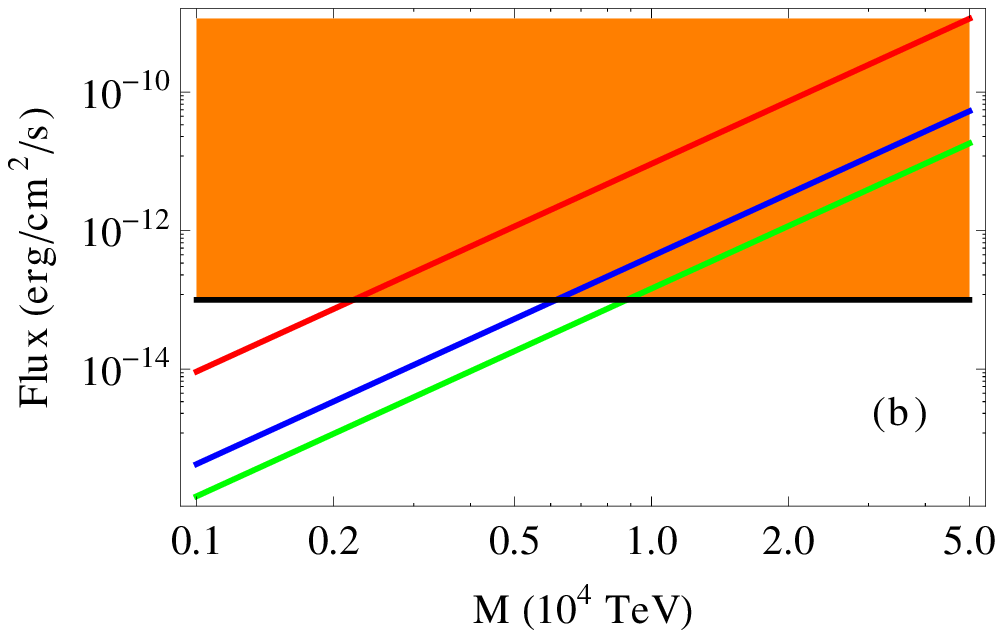}
\includegraphics[height=2.5in,width=2.5in]{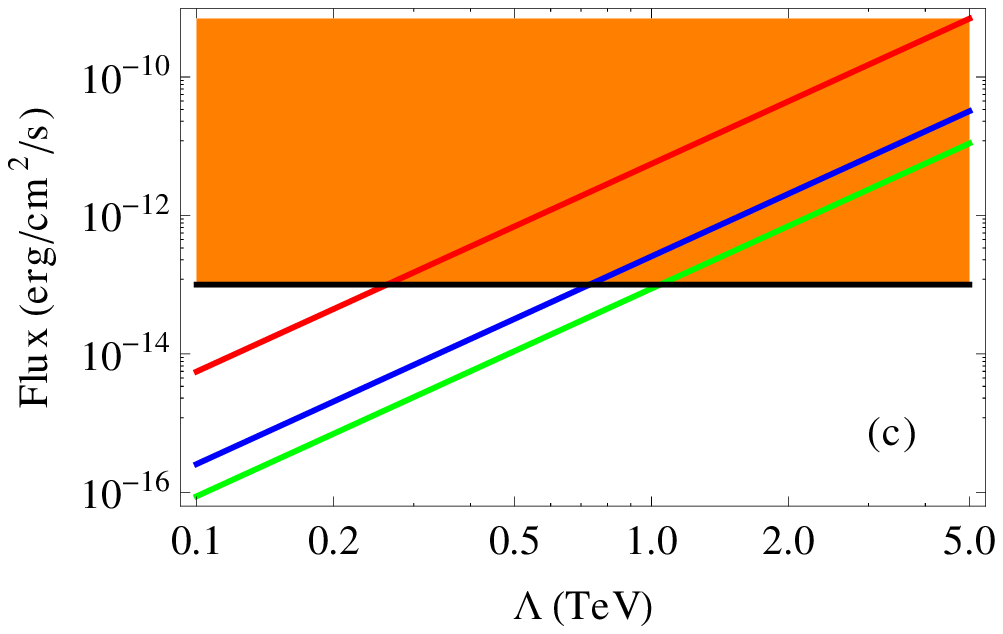}
\end{center}
\caption{Constraints on the parameters $M,\Lambda$ due to
the observations made by {\it Chandra} telescope. In plot (a) anything
below the thick red line is excluded in the plane $M-\Lambda$.
The blue and magenta lines are contour lines where the neutrino
mass scales are 1 eV and 0.01 eV, respectively.
Points along the green line in this plot fits the 100$\%$ relic
density of the present universe. The shaded yellow
region is allowed by both the X-ray flux and the neutrino mass scale
restrictions.
In plots (b) and (c) X-ray flux from various clusters are plotted
as a function of $M$ and $\Lambda$. $\Omega_sh^2$ = 0.1
is taken in both these plots.
The red, blue and green lines in
these plots are for the X-ray flux due to sterile neutrinos coming from
the clusters Virgo center, NGC 3198, NGC 4123,
respectively The shaded orange region in these plots is excluded.}
\label{F:VC}
\end{figure}

The constraints in the plane $M-\Lambda$ due to Virgo cluster as
observed by {\it Chandra} telescope is given
in Fig. \ref{F:VC}(a).
We have used $M^{\rm fov}_{\rm DM}=10^{13}M_\odot$
and $D_L=$ 20.7 Mpc \cite{AFT} in eq. (\ref{E:cdmF}) and demanded
that the flux should be less than the detectable limit $10^{-13}$ erg
cm$^{-2}$ s$^{-1}$. The thick red line in this figure represents
this flux restriction. Anything below this line can be considered
to be ruled out. The blue and magenta lines in this plot represent
contour lines of constant neutrino mass scale of 1 eV and 0.01 eV,
respectively. The first relation of eq. (\ref{E:masnmix}) gives
neutrino mass scale as a function of $\Lambda$ and $M$. Although
the neutrino mass scale is around 0.1 eV, in Fig. \ref{F:VC}(a)
we have allowed flexibility in the neutrino mass scale to be between
1 eV and 0.01 eV. The solar neutrino mass scale is $\approx$
0.009 eV, which is a few factor less than the atmospheric scale
of $\approx$ 0.05 eV. It is stated before that the neutrino mass
scale of eq. (\ref{E:masnmix}) gives a rough estimation.
As a consequence of this it may not be realistic
to fit the neutrino
oscillation data for a neutrino mass scale less than about 0.01 eV.
Hence, the yellow shaded region in Fig. \ref{F:VC}(a) can be taken
to be as allowed region by the X-ray flux restriction and the
neutrino mass of this model.
However, in the
flux restriction we do not demand that the sterile neutrinos to
be exactly as warm dark matter particles, but we only require them to
decay radiatively to a photon. Stringent bounds can be obtained
on these parameters if we demand the sterile neutrino to fit
100$\%$ of the current relic density of the universe. For relic
abundance we have taken $\Omega_sh^2=$ 0.1 which falls within
the experimental limits of it as observed by the Wilkinson Microwave
Anisotropy Probe (WMAP) \cite{wmap}. The thick
green line in Fig. \ref{F:VC}(a) represents a contour line
in the parameters where $\Omega_sh^2=$ 0.1.
We have given the individual constraints
on the $M$ and $\Lambda$ in Figs. \ref{F:VC}(b) and \ref{F:VC}(c),
which arise due to both the X-ray flux restriction and also
by the demand of 100$\%$ relic abundance from the sterile neutrino.
By putting $\Omega_sh^2=$ 0.1,
we have plotted flux versus $M$ in Fig. \ref{F:VC}(b) and flux
versus $\Lambda$ in Fig. \ref{F:VC}(c), using eqs. (\ref{E:cdmF}) and
(\ref{E:cdmRD}). In both these plots
we have also given restrictions coming from clusters
NGC 3198 and NGC 4123, in order to compare with the results obtained from
Virgo cluster. The red line in both these plots is for Virgo
cluster, the blue line is for NGC 3198 and the green line is for NGC 4123, respectively.
The values of $(M^{\rm fov}_{\rm DM},D_L)$ for NGC 3198 and NGC 4123
are (3.62$\times 10^{11}M_\odot$, 18.34 Mpc) and (1.85$\times 10^{11}M_\odot$,
22.4 Mpc), respectively \cite{AFT}. The orange shaded region in both
Figs. \ref{F:VC}(b)
and \ref{F:VC}(c) is the disallowed region. We can see that Virgo cluster
put upper bound on the $M$ to be less than about 0.2$\times 10^4$ TeV and the corresponding
bound on the scale $\Lambda$ is less than 0.3 TeV. The bounds
from the clusters of NGC's on $M$ and $\Lambda$ are reasonable but they
are less stringent than that due to the Virgo cluster.

We remind here that we have neglected the ${\cal O}(1)$ constants in
eq. (\ref{E:masnmix}). These ${\cal O}(1)$ constants will change the
flux and relic density relations, eqs. (\ref{E:cdmF}) and (\ref{E:cdmRD}),
by some factors. If the resultant
factor in the flux relation of eq. (\ref{E:cdmF}) is less than one, we expect
the red line of Fig. \ref{F:VC}(a) to shift little bit down, and hence the constraint
due to flux restriction will be weakened. The opposite effect happens if this
factor is greater than one. Similar behavior can be expected for the
relic density constraint of eq. (\ref{E:cdmRD}) due to these factors.
Although we do not expect the constraints to change significantly
away from what we have obtained in Fig. \ref{F:VC}, we leave this for a detailed
study later. Another point to note here is that in Fig. \ref{F:VC}(a),
we have plotted from $\Lambda$ = 0.1 TeV, where the corresponding allowed value
of $M$ by neutrino mass and flux restriction would be $\sim 10^3$ TeV.
For this $\Lambda$ and $M$ values we not only get 0.1 eV neutrino masses
but also viable keV sterile neutrinos, as explained below eq. (\ref{E:masnmix}).
Although big-bang nucleosynthesis put
a lower bound on $\Lambda$ to be $\sim$ 1 GeV \cite{dCDN}, for such a low value of
$\Lambda$ it is not possible to get $m_\nu\sim$ 0.1 eV and $m_s\sim$ keV by
using eq. (\ref{E:masnmix}).

After giving constraints on the composite Dirac neutrino model
due to the observations made by {\it Chandra} telescope, we now give constraints
on this model from different data analysis.
In \cite{BNRS}, data collected by {\it XMM-Newton} on Coma and Virgo clusters
have been analyzed. Here, it has been pointed that the background due
to continuum X-rays from intra-cluster gas can be reduced compared
to the signal if the flux at the periphery is collected. Accordingly, we
can analyze flux from the center and periphery of both Coma and Virgo
clusters. From this analysis we can see that the data from Coma periphery put
stringent limits among all these observations. The method employed
in getting the constraints is explained as follows. The experiment
{\it XMM-Newton} has collected
data in the energy region of 0.5 to 8.5 keV. This region has been
divided into bins of size 0.2 keV. By demanding the X-ray flux from dark matter in the
center of each of the energy bin to be less than the detected flux
in that bin, an exclusion plot in the plane $\sin^22\theta-m_s$ can
be obtained. Below we give the formulas for the flux with projected
radius $r$ from Coma and Virgo clusters, respectively \cite{BNRS}:
\begin{eqnarray}
F_{\rm Coma}&\approx &6.7\times 10^{-8}\theta^2\left(\frac{m_s}{1{\rm keV}}\right)^5
g(r)~{\rm erg~ cm^{-2}~ s^{-1}},
\nonumber \\
F_{\rm Virgo}&\approx &7.7\times 10^{-9}\theta^2\left(\frac{m_s}{1{\rm keV}}\right)^5
g(r)~{\rm erg~ cm^{-2}~ s^{-1}}.
\label{E:bnrsF}
\end{eqnarray}
Here, $g(r)$ is a geometric factor with projected radius $r$. For
Coma center and periphery its values are 0.82 and 0.18 respectively,
whereas, for Virgo center (M87) the geometric factor is 6.8 \cite{BNRS}.
Although the actual exclusion line in the plane $\sin^22\theta-m_s$
would be obtained by demanding the flux values from above equations
in a energy bin to be less than the measured flux in that energy bin,
a rough understanding on these constraints can be understood by plotting
the contour lines in the plane $\sin^22\theta-m_s$ for a constant flux at
0.5 keV and 8.5 keV. Imagine we are doing analysis on the
data from Coma periphery. By putting the geometric factor for
this source in $F_{\rm Coma}$, we plot contour lines in $\sin^22\theta-m_s$
for the relations: $F_{\rm Coma}=F_{0.5keV}$ and $F_{\rm Coma}=F_{8.5keV}$,
where $F_{0.5keV}$ and $F_{8.5keV}$ are the observed flux
values in the experiment at the energy bins containing 0.5 keV and
8.5 keV, respectively. These two contour lines give a rough estimation
on the exclusion area in the plane $\sin^22\theta-m_s$. But the
actual exclusion line from the full analysis lie in between these
two extreme contour lines. We demonstrate this when we present our
results on the composite Dirac neutrinos below. However,
the exclusion in the parameters of $(\sin^22\theta,m_s)$ from the
analysis done on the coma periphery by the {\it XMM-Newton} telescope
can be parameterized as \cite{BNRS}
\begin{equation}
\sin^2(2\theta)\leq 3.76\times 10^{-4}\left(\frac{m_s}{\rm 1keV}\right)^{-5.63}
\label{E:empf}
\end{equation}

To translate the above restrictions from the data by {\it XMM-Newton} into
the composite Dirac neutrino model, we include a half-factor in
the flux relations since in the analysis described above Majorana
neutrinos have been assumed. Using eq. (\ref{E:masnmix}), the flux
relations for the Dirac neutrinos turn out to be:
\begin{eqnarray}
F_{\rm Coma}&\approx &3.35\times 10^{-11}\left(\frac{v}{\rm 1TeV}\right)^2
\left(\frac{\Lambda}{\rm 1TeV}\right)^{15}
\left(\frac{10^4~\rm TeV}{M}\right)^{12}g(r)~{\rm erg~ cm^{-2}~ s^{-1}},
\nonumber \\
F_{\rm Virgo}&\approx &3.85\times 10^{-12}\left(\frac{v}{\rm 1TeV}\right)^2
\left(\frac{\Lambda}{\rm 1TeV}\right)^{15}
\left(\frac{10^4~\rm TeV}{M}\right)^{12}g(r)~{\rm erg~ cm^{-2}~ s^{-1}}.
\label{E:CnV}
\end{eqnarray}
The relation of eq. (\ref{E:empf}) will translate in our case as
\begin{equation}
\left(\frac{M}{10^4~\rm TeV}\right)^{13.26}\geq\frac{2}{3.76}10^{1.63}
\left(\frac{v}{\rm 1TeV}\right)^2\left(\frac{\Lambda}{\rm 1TeV}\right)^{16.89}.
\label{E:empMod}
\end{equation}

\begin{figure}
\begin{center}
\includegraphics[height=2.5in,width=2.5in]{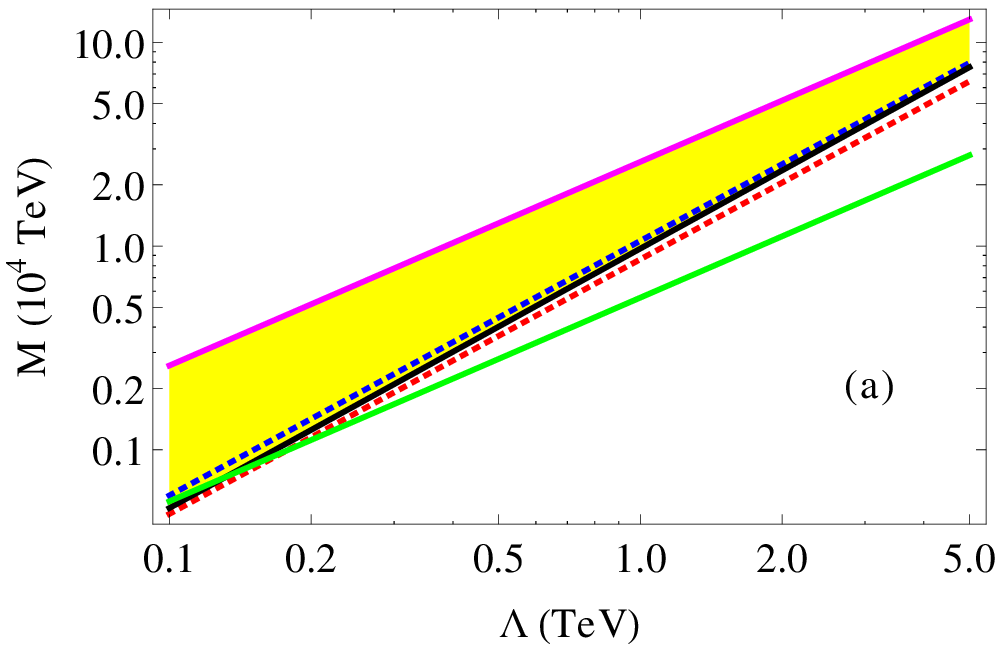}
\includegraphics[height=2.5in,width=2.5in]{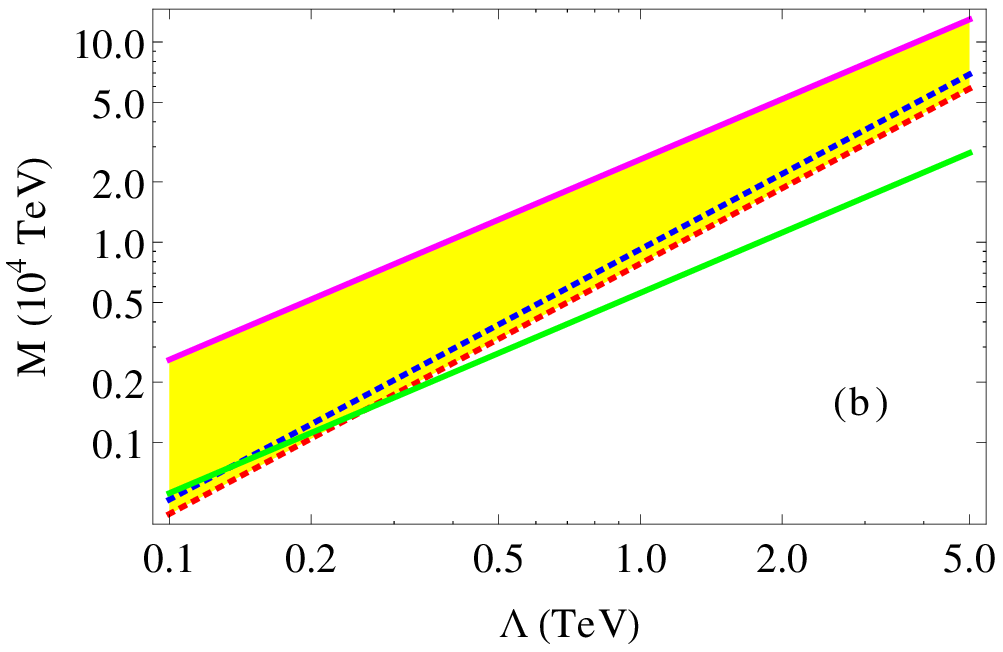}

\includegraphics[height=2.5in,width=2.5in]{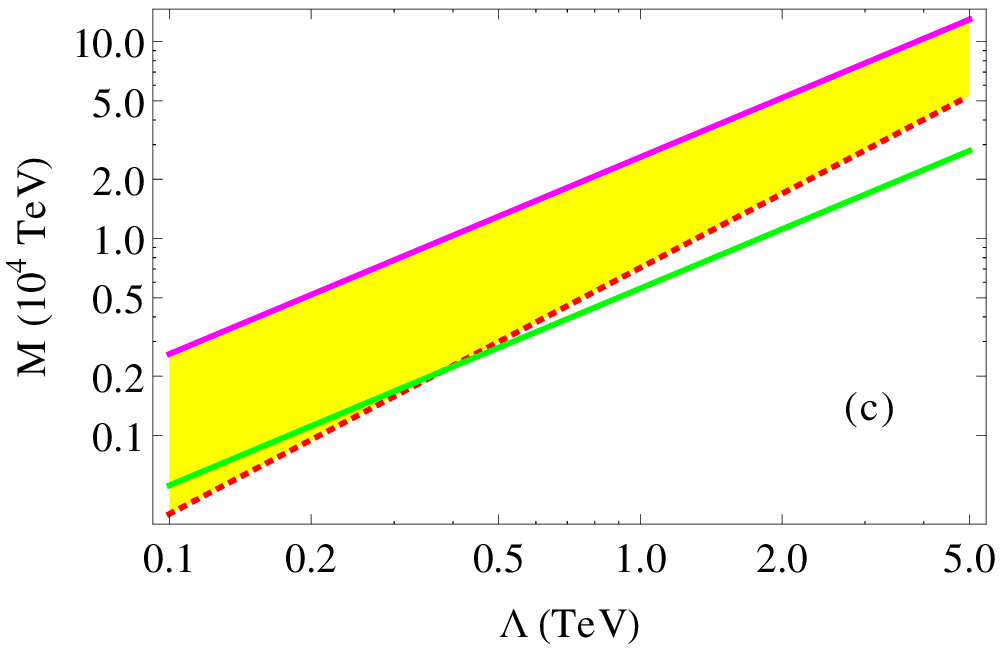}
\end{center}
\caption{Constraints in the plane $M-\Lambda$ from the data
collected by {\it XMM-Newton} telescope. Anything below the red dotted
line in these figures are ruled out. The green and magenta lines
are contour lines where the neutrino mass scale are 1 eV and
0.01 eV respectively. The yellow shaded region is allowed by
both the X-ray flux and neutrino mass scale restrictions.
Constraints in the plots (a), (b) and (c)
are due to the data from Coma periphery, Coma center and Virgo
center, respectively.}
\label{F:XMM}
\end{figure}
The constraints due to Coma periphery are given in Fig. \ref{F:XMM}(a).
The red and blue dotted lines in this plot represents the constant
contour lines for $F_{\rm Coma}=F_{0.5 \rm keV}$ and $F_{\rm Coma}=F_{8.5 \rm keV}$,
respectively. The actual constraint due to the full analysis of
eq. (\ref{E:empMod}) is given in thick dark line which almost touches
the constant contour lines at both the extreme ends. The values for
the flux from Coma periphery are taken as: $F_{0.5 \rm keV} = 1.1
\times 10^{-12}~{\rm erg~cm^{-2}~s^{-1}}$, $F_{8.5 \rm keV} = 9.0
\times 10^{-14}~{\rm erg~cm^{-2}~s^{-1}}$ \cite{BNRS}. Anything
below the dark line is disallowed. But it can be noticed that the
red dotted line in this plot gives a conservative bound and, whereas
the blue dotted line gives a slightly stringent overestimated bound compared to
the dark line. Hence the constant flux contour lines at the extreme
energy bins give a rough estimation of the constraints. In Fig.
\ref{F:XMM}(a) we have also given the contour lines of constant
neutrino mass scale. The green and magenta lines in this plot represent
a constant neutrino mass scale of 1 eV and 0.01 eV respectively. As explained
previously that the neutrino mass scale below 0.01 eV may not fit the
neutrino oscillation data, and hence the yellow shaded region between the
dark line and the magenta line is considered to be allowed region
in the plane $M-\Lambda$.
In Fig. \ref{F:XMM}(b) constraints due to
the Coma center are given in the plane $M-\Lambda$.
We have given the constant flux contour lines of $F_{\rm Coma}=F_{0.5 \rm keV}
=1.6\times 10^{-11}~{\rm erg~cm^{-2}~s^{-1}}$
and $F_{\rm Coma}=F_{8.5 \rm keV}=2.2\times 10^{-12}~{\rm erg~cm^{-2}~s^{-1}}$
\cite{BNRS}, which are indicated in red and
blue dotted lines, respectively. Anything below the red dotted line
can be considered to be excluded, but the blue dotted line gives a
somewhat overestimated stringent bound. The actual constraint from
the full analysis would lie somewhere between the red and blue dotted
lines, similar to the case of Coma periphery which is explained above.
The green and magenta lines in this plot give constant neutrino
mass scale of 1 eV and 0.01 eV respectively. The yellow shaded region
is allowed from the X-ray flux and the neutrino mass constraints.
In Fig. \ref{F:XMM}(c) we
have given constraints due to Virgo center as observed by the
{\it XMM-Newton} telescope. In this plot we have a constant contour
line of $F_{\rm Virgo}=F_{1 \rm keV}=4.8\times 10^{-11}~{\rm erg~cm^{-2}~s^{-1}}$
\cite{BNRS}, which is shown in the red dotted line. Like for the
previous plots anything below this line can be considered to be
ruled out, and also the yellow shaded region is allowed from the
X-ray flux and the neutrino mass scale constraints.

Comparing the constraints from various sources in Fig. \ref{F:XMM}, Coma
periphery put stringent limits due to the fact that the observed
flux from it is at least an order less than that due to the
other sources. In any of the plots of Fig. \ref{F:XMM} we have not
applied the restriction that the sterile neutrino should fit the
100$\%$ of the relic density of the universe. However, if applied,
like we have done in Fig. \ref{F:VC}, we may expect similarly stringent
limits on both the scales $M$ and $\Lambda$.

\section{Conclusions}
\label{S:con}

In the era of the Large Hadron Collider, where we are in now, it is important
to analyze the new physics models and if possible put constraints
in a model. However, apart from collider data, astrophysical
experiments also play a role in probing the new physics models
which have relation to cosmology. We have analyzed one of those
models where Dirac neutrino masses can only appear as a result of
confinement of a UV preonic theory at a high scale $M$ into some
hidden sector and standard model fields, and the hidden sector confines
into right-handed neutrinos at a lower scale $\Lambda\ll M$ \cite{GR}.
Both the scales $M$ and $\Lambda$ determine the masses
of the active and sterile neutrinos, and also the mixing angle
between them. In the case of secondary mass generation for the
sterile neutrinos, and for $\Lambda\sim$ TeV and $M\sim 10^4$ TeV,
we get keV mass sterile neutrinos with a mixing
angle of $\sim 10^{-5}$ with the active neutrinos, apart from getting
the correct neutrino mass scale of 0.1 eV. The keV
mass sterile neutrinos form viable warm dark matter candidates.
One of the channels to probe the sterile
neutrino is its decay to an X-ray photon and an active neutrino, which
has been looked in astrophysical experiments like {\it Chandra} and
{\it XMM-Newton}. The mass scale and the mixing angle for
the sterile neutrino in this particular model falls in the right ball park
region analyzed in these experiments. Hence, we have studied
constraints on the model parameters $\Lambda$ and $M$ due to
the negative search of these particles in the astrophysical
experiments. A previous study on this kind of model has put a lower
bound of $\sim$ 1 GeV on the parameter $\Lambda$
from the big-bang nucleosynthesis \cite{dCDN}. Whereas, in this
work we have shown that from the X-ray analysis and also if
the sterile neutrino make up 100$\%$ of the dark matter,
we can get upper bounds on both of the parameters $\Lambda$ and $M$.

More specifically, from the analysis of the data collected by
{\it Chandra} telescope on Virgo cluster we have obtained an upper bound on
the scale $M$ to be about 2$\times 10^3$ TeV if it
fits the 100$\%$ of the current relic abundance of
the universe. Under the same assumption the corresponding
upper bound on the scale $\Lambda$ is $\sim$ 300 GeV. From the data
collected by the {\it XMM-Newton} telescope on the periphery of
the Coma cluster, we got stringent limits in the plane
of $M$ and $\Lambda$. However,
in this analysis we have not applied the condition that
the sterile neutrino should fit the 100$\%$ of the
relic abundance of the universe, but, if applied we may
get similar constraints as mentioned above. In both the
analysis that we have mentioned above, we have also given
region of parametric space in the plane $M-\Lambda$ allowed
by both the neutrino mass scale and the X-ray flux restrictions.

\end{document}